\documentclass[%
preprint,       % for two-column layout like published article
superscriptaddress, % for author affiliations as superscripts
amsmath,amssymb,  % common packages for equations
aps,           % style: 'aps', 'aip', 'prl', etc.
prb            % journal style: prl, pra, prb, rmp, etc.
]{revtex4-2}
\newcommand{\ta}{\tilde{A}}
\newcommand{\bta}{\tilde{\mathbf{A}}}
\newcommand{\ros}{R_{\text{os}}}

\newcommand{\vinn}{\tilde{v}_{\text{in},n}}

\newcommand{\voutn}{v_{\text{out},n}}

\newcommand{\vout}{\mathbf{v}_{\text{out}}}
\newcommand{\vin}{\tilde{\mathbf{v}}_{\text{in}}}

\newcommand{\nin}{N_{\text{in}}}
\newcommand{\nout}{N_{\text{out}}}
\newcommand{\ber}{\text{BER}}
\usepackage{tensor}
\usepackage{cleveref}
\usepackage{booktabs}
\usepackage{graphicx}
\usepackage{fancyhdr}
\begin{document}
	\author{Nicholas Cox}
	\email{nicholas.a.cox34.civ@us.navy.mil}
	\author{Joseph Murray}
	\author{Ross T. Schermer}
	\affiliation{U.S. Naval Research Laboratory, 4555 Overlook Ave, SW, Washington, DC 20375, USA}
	\author{Shuo S. Pang}
	\affiliation{Mileva, Oviedo FL 32765}
	\author{Christopher Long}
	\author{Gajadhar Joshi}
	\author{Nicholas Nobile}
	\author{Raktim Sarma}
	\affiliation{Center for Integrated Nanotechnologies, Sandia National Laboratories, 1515 Eubank Blvd SE, Albuquerque, NM 87185, USA}

	\author{Brandon Redding}
	\affiliation{U.S. Naval Research Laboratory, 4555 Overlook Ave, SW, Washington, DC 20375, USA}

	\title{Nonlinear pluggable optics: Digital signal processing-free Intensity Modulated Direct Detection links using analog photonic Next Generation Reservoir Computing }

	\begin{abstract}
		In this work, we propose a Nonlinear Pluggable Optic (NLPO) transceiver that combines the low latency and low power consumption of Linear Pluggable Optics (LPO) with the range and robustness of digital signal processing (DSP)-based transceivers. The proposed NLPO uses an analog photonic Next-Generation Reservoir Computing (NGRC) architecture, constructed on a photonic integrated circuit (PIC), to compensate for electrical-domain distortions as well as optical-channel impairments from chromatic dispersion and Kerr nonlinearity. Focusing on a simulated 50 GBd PAM-4 link, we find that the NGRC-based NLPO not only extends the range of LPO, but actually outperforms DSP-based solutions as well. Our simulations reveal two key advantages compared to DSP-based Intensity Modulation/Direct Detection (IM/DD) links: (1) the NGRC can take advantage of the optical phase information without requiring a local oscillator and (2) the NGRC can optically sample the transmitted data well above the symbol rate without requiring high-bandwidth electronics. This work showcases the potential for photonic NGRCs to outperform state-of-the-art digital solutions in real-world applications and opens a path to low-latency, lower-power IM/DD links at ranges of 10s of km.
	\end{abstract}
	
	\maketitle
	\thispagestyle{fancy}
	\section{Introduction}
	
	The escalating bandwidth requirements for high-speed data center interconnects to support AI/ML clusters have led to a surge in the adoption of digital signal processing (DSP)-free linear pluggable optic (LPO) transceivers due to their reduced power consumption and latency. However, LPO architectures lack the ability to correct for nonlinear distortions, leaving them vulnerable to impairments from non-ideal encoding electronics as well as the optical-domain distortions introduced by chromatic dispersion and the Kerr effect. The encoding limitation imposes strict constraints on the LPO transceiver components to ensure linear response of all modulators and electrical amplifiers, while optical propagation effects restrict the maximum reach and limit operation to the O-band where dispersion is minimized. The challenges associated with LPO become more pronounced at higher symbol rates, for which transmitted signals are increasingly susceptible to nonlinearities. Here, we propose a DSP-free nonlinear pluggable optic (NLPO) that uses an analog photonic next-generation reservoir computer to overcome the limits of LPO while maintaining their advantages of low power consumption and latency.
	
	Next-generation reservoir computing (NGRC) is a machine learning architecture designed to perform inference on time series data. NGRC simplifies the standard reservoir computing (RC) paradigm  by removing the need for recurrent feedback, relying instead on the generation of feature vectors formed by applying linear and nonlinear transformations to a set of discrete time samples of the data under investigation. While the memory of conventional RCs depends on the recurrent feedback structure of the reservoir \cite{jaegerHarnessingNonlinearityPredicting2004,appeltantInformationProcessingUsing2011,duportAllopticalReservoirComputing2012,paquotOptoelectronicReservoirComputing2012}, NGRC enables explicit control over the system memory to match the needs of a given problem.  
	
	NGRCs were first introduced in 2021 as a digital, software-based machine learning technique with better interpretability and easier hyperparameter optimization than traditional RCs \cite{gauthierNextGenerationReservoir2021}. In 2024, researchers showed that NGRCs could be implemented using analog photonics with the potential for lower power consumption and reduced latency \cite{coxPhotonicNextgenerationReservoir2024,wangUltrafastSiliconPhotonic2024,wangOpticalNextGeneration2025}. Photonic NGRCs are particularly promising for applications requiring real-time inference at speeds exceeding digital clock rates, such as channel equalization and distortion compensation in communication links \cite{duportAllopticalReservoirComputing2012,duportFullyAnaloguePhotonic2016,antonikOnlineTrainingOptoElectronic2017,argyrisPhotonicMachineLearning2018,argyrisPAM4Transmission15502019,argyrisComparisonPhotonicReservoir2020,coxPhotonicFrequencyMultiplexed2025,coxPhotonicNextgenerationReservoir2025,wangTerabitIntegratedNeuromorphic2025}. 
	
	Today, distortion compensation in fiber optic communication links relies heavily on digital signal processing (DSP). However, as data rates continue to increase, relying on DSP becomes difficult: DSP already consumes 50\% of the power in transceiver modules \cite{nagarajanLowPowerDSPBased2021} and introduces unacceptable latency for many applications, including AI training \cite{wangTerabitIntegratedNeuromorphic2025,marvellMarvellAraPAM42024}. In addition, the operating range of widely deployed, low-cost intensity modulated direct detection (IM/DD) links decreases with data rate: since IM/DD links only measure intensity, phase information is lost and DSP can only compensate for modest levels of distortion. Coherent modulation formats support longer range by measuring the optical phase, but they require a local oscillator and a frequency stabilized source laser, leading to prohibitively high cost for dense data-center and metropolitan links \cite{huang800GbTransmission2025}. Recent efforts at LO-free coherent communication schemes have attempted to bridge the gap between low-cost IM/DD and the coherent links used in long-haul communications, but these approaches transmit a carrier along with the data which reduces spectral efficiency \cite{dingFullduplexBroadcastRoFWDMPON2018} or requires additional measurement channels and complex DSP \cite{huang800GbTransmission2025}. 
	
	In the past few years, the power consumption and latency pitfalls of DSP led researchers to introduce linear receive optics (LRO) and linear pluggable optics (LRO), which remove DSP modules from the receive side or both sides, respectively, of IM/DD links. However, these systems typically require customized electronic front-ends to ensure linear data encoding, are constrained to the near-zero dispersion regime around 1310 nm, and are limited to shorter range than DSP-based IM/DD links\iffalse[ref]\fi. 
	
	In this work, we show that a photonic NGRC is particularly well suited for distortion compensation in IM/DD links, providing a path to a DSP-free Nonlinear Pluggable Optic that combines the low latency and low power consumption of LPO with longer range operation and the ability to operate across the C-band. Focusing on the widely used 50 GBd PAM-4 modulation format (transmitting 100 Gb/s), we show that photonic NGRCs can not only extend the range of LPOs, but actually outperform DSP-based IMDD links, extending the range from ~1 km to $>$50 km while maintaining a bit error rate (BER) below the KP4 RS(544,514) FEC threshold of $2.2\times 10^{-4}$ \cite{patraUpdatesConcatenatedFEC2022}. To achieve this performance, we optimized the photonic NGRC architecture and found that there are two factors that enable the photonic NGRC to outperform DSP-based receivers: (1) the photonic NGRC operates on the complex electric field, providing access to phase information that is lost in a DSP-based approach and (2) the photonic NGRC can oversample the data at rates far exceeding the symbol rate without requiring high-bandwidth electronics. Crucially, we show that the photonic NGRC does not require a local oscillator to take advantage of transmitted phase information. The optimized NGRC design virtually oversamples the transmitted data at rates up to 250 GS/s, which would be prohibitively expensive in a digital implementation. To the best of our knowledge, this is the first work to show that oversampling can be a major advantage of a photonic reservoir computer. 
	
	Here, we present a photonic integrated circuit (PIC)-based NGRC architecture and optimize its design to achieve the best equalization results. While this work is simulation-based, these optimizations culminate in a performance test including the effects of amplified spontaneous emission (ASE) and shot noise, demonstrating the architecture's robustness against real-world impairments. To this end, we demonstrate the tuning of photonic NGRC hyperparameters toward a target application by adjusting the feature vector size and optimizing the memory and oversampling rate. This work identifies the key advantages of analog photonic NGRCs and highlights their potential to outperform state-of-the-art digital solutions in real-world applications.

	\section{Digital vs. Photonic NGRCs}
	Digital NGRCs operate by generating a feature vector from explicit polynomial combinations of the input data over a fixed number of time steps \cite{gauthierNextGenerationReservoir2021,gauthierLearningUnseenCoexisting2022}. This feature vector is treated as the equivalent of the reservoir state in a traditional reservoir computer: the only training required is a simple regression step which applies weights to the feature vector to determine the target output. The NGRC also benefits from intuitive and easily optimized hyperparameters, consisting of the system memory and the polynomial order (linear, quadratic, cubic, etc.) that are included in the feature vector. The downside of NGRCs is that the size of the feature vector grows rapidly with memory length and polynomial order, and must be recomputed at each time step.
	
	Photonic NGRCs have the potential to accelerate this process by using high-bandwidth analog photonic components to generate the feature vector. If the trained weights are also applied in the photonic domain, the entire NGRC process, from feature vector formation to weighting and inference can be performed at speeds exceeding digital clock rates \cite{coxPhotonicFrequencyMultiplexed2025,wangUltrafastSiliconPhotonic2024,wangTerabitIntegratedNeuromorphic2025}. To date, photonic NGRCs have been proposed based on temporal multiplexing in optical fiber \cite{coxPhotonicNextgenerationReservoir2024}, frequency multiplexing using optical frequency combs \cite{coxPhotonicFrequencyMultiplexed2025}, and spatial multiplexing in free-space using spatial light modulators \cite{wangOpticalNextGeneration2025} or in waveguides on a photonic integrated circuit (PIC) \cite{wangUltrafastSiliconPhotonic2024,wangTerabitIntegratedNeuromorphic2025}. The spatially-multiplexed PIC-based implementation is particularly promising for high-bandwidth distortion compensation in fiber communication applications due to its potential for integration with existing PIC-based transceivers and low power consumption due to the passive generation of the feature vector \cite{wangTerabitIntegratedNeuromorphic2025}. As a result, this work focuses on a spatially-multiplexed photonic NGRC suitable for implementation on a PIC. 
	
	From an algorithm perspective, there is one major distinction between photonic and digital NGRCs. In the original software-based digital NGRC formulation, the feature vector is formed by explicitly computing each polynomial combination of the input data up to the desired order. This can be challenging to implement in an analog photonic platform, since it requires a variety of nonlinear operations. Instead, many of the photonic NGRCs rely on an elegant work-around in which the optically encoded input data is coherently mixed to form a speckle pattern and this speckle pattern is used as the feature vector \cite{coxPhotonicNextgenerationReservoir2024,wangOpticalNextGeneration2025,wangUltrafastSiliconPhotonic2024, wangParallelOpticalNextgeneration2026}. The speckle pattern can be decomposed into polynomial combinations of the input data, providing access to the same terms as a second order digital NGRC. To add linear terms to the feature vector, a local oscillator can be included \cite{wangUltrafastSiliconPhotonic2024}. The only constraint on these speckle-based feature vectors is that, for an input vector of length $\nin$, the feature vector must contain $\sim \nin^2$ elements to ensure equivalence to a digital NGRC feature vector (see Appendix \ref{sec:ngrc_model} for details). However, as we will show, a feature vector with fewer elements can still provide excellent performance. Interestingly, some digital NGRCs have begun using this speckle-based approach because it allows for more flexibility in selecting feature vector size, making it possible to reduce the computational load without significantly impacting performance \cite{cestnikNextgenerationReservoirComputing2026}.
	
	\section{PIC-based Photonic NGRC Architecture for Channel Equalization}\label{sec:pic}
	
\begin{figure}[h]
		\centering
		\includegraphics{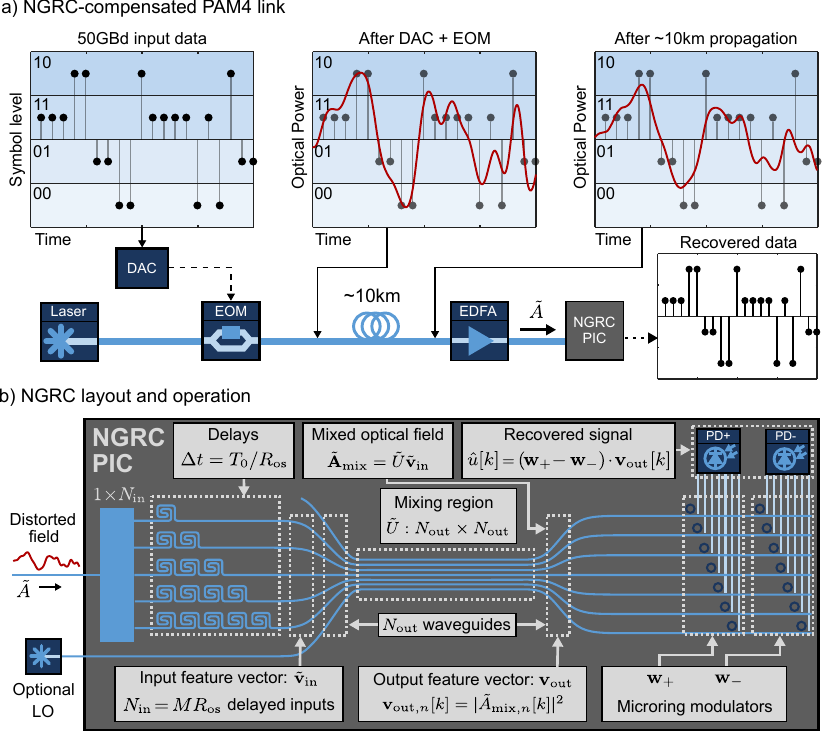}
		\caption{a) Demonstration of the distortions present in a 50Gbaud PAM4 link. b) Diagram of the photonic NGRC PIC which compensates for channel distortions in the analog optical domain.}\label{fig:diagram}
	\end{figure}

	An overview of the photonic NGRC design is shown in \cref{fig:diagram}. In this work, we consider 50 GBd PAM4 data which is distorted at the encoding stage by the digital to analog converter (DAC) and Mach-Zehnder modulator and during propagation in the fiber by dispersion and Kerr nonlinearity. \Cref{fig:diagram}a provides an example of the distortions by showing the simulated optical power after the DAC and modulator and again after propagating through a 10 km fiber. A key difference between these effects is that the encoding process distorts the optical power, whereas optical propagation directly affects the complex optical field, which cannot be directly measured. Details on the modeled link are provided in Appendix \ref{sec:dis}.
	
	In a conventional PAM-4 link, the distorted signal in \cref{fig:diagram}a would be recorded at the data rate (50 GS/s) and processed using DSP to recover the transmitted symbols. Moderate signal impairment due to the DAC can usually be addressed because its contribution to the measured optical power is described by a convolution with a linear finite impulse response (FIR) filter. The role of the DSP is then to approximate and apply the corresponding inverse filter to compensate for this effect. It is also straightforward to mitigate the signal degradation imposed by the idealized nonlinear voltage-to-optical power response function of the modulator; this distortion is pre-compensated by adjusting the voltage levels used to encode symbols. However, dispersion and Kerr nonlinearities encountered during optical propagation prevent DSP from performing satisfactory symbol recovery for 1550 nm links beyond $\sim$ 1 km \cite{eiseltEvaluationRealTime82017}. This failure occurs because optical propagation distorts the complex field, and square-law photodetection removes the phase information needed to characterize those distortions.
	
	The optical propagation-induced reduction in equalization performance is especially pronounced in LPO-based systems. Since LPO architectures rely on SerDes-stage ASICs for compensation, they lack the flexibility of DSP to isolate and correct individual distortion sources. Attempting to simultaneously equalize mixed optical and electronic distortions in the digital domain is highly complex, which typically forces these systems to rely on optical components with strict linearity requirements. In contrast, we show that the NGRC-based NLPO architecture can tolerate significant distortions from non-ideal DACs and modulators, permitting the use of more affordable components. Furthermore, leveraging this accessible phase data allows the system to successfully equalize both optical and electronic distortions in a single step.
	
	A schematic of the PIC-based NGRC is shown in \cref{fig:diagram}b. The PIC is designed to implement the entire NGRC process by: (1) creating an input feature vector consisting of the transmitted data over $\nin$  time steps, (2) mixing the input feature vector to create an output feature vector, (3) applying trained weights to the output feature vector, and (4) detecting the weighted output to recover the original symbols at the encoding data rate. Here, we describe how to implement each step on a PIC. The transmitted light is first coupled from the optical fiber to a single-mode waveguide on the PIC. A fiber amplifier can be inserted before the NGRC device to compensate for losses due to propagation through the link and PIC. The input feature vector, $\vin$, is generated on the PIC by dividing the light into $\nin$ paths and adding a delay of length $n\Delta L$ to the $n^{\text{th}}$ path, where $\Delta L= (T_0/\ros)c/n_g$, $T_0$ is the symbol period ($T_0=20$ ps for the 50 GBd data rate considered here), $\ros$ is the oversampling rate of the NGRC, $c$ is the speed of light, and $n_g$ is the waveguide group index. Since the input data is oversampled by an integer factor $\ros$, the NGRC memory in terms of the number of symbol periods represented in the input feature vector is given by $M=\nin/\ros$. The input feature vector $\vin$ is denoted with a tilde to indicate that it is a complex vector with elements equal to the optical field including phase information. Note that an optional LO could be included in the formation of the input feature vector. However, we will show that an LO is not necessary for the PAM4 channel equalization task considered in this work. 
	
	The input feature vector is then converted into a speckle-pattern based output feature vector, $\vout$, by coherently mixing each of the input feature vector elements to form a number of outputs $\nout$ (where $\nout > \nin$). Note that, in contrast to the input feature vector, the elements $\voutn$ of the output feature vector $\vout$ correspond to the real-valued optical power of each output. In general, a variety of passive, linear photonic devices could be used to create the speckle pattern, including an on-chip scattering region \cite{wangIntegratedPhotonicEncoder2024}, a star coupler \cite{wangUltrafastSiliconPhotonic2024}, or, as shown in \cref{fig:diagram}b, an array of evanescently coupled waveguides (see Appendix \ref{sec:ngrc_model} for more detail and simulations of the evanescently coupled waveguide mixing region). After the mixing region, the $\nout$ output waveguides containing each of the $\nout$ elements in the output feature vector are separated to prevent additional coupling. In this work, we assume the mixing region can be described by a random, $\nout \times \nout$ unitary transmission matrix $\tilde{U}$ that applies to an $\nout$-dimensional vector formed by embedding the $\nin$-dimensional input feature vector into an $\nout$-dimensional space. Appendix \ref{sec:ngrc_model} provides a full mathematical model for the input and output feature vectors and shows how this speckle-based output feature vector can be decomposed into the polynomial terms of a certain form of second-order digital NGRC. 
	
	To apply the trained weights, we can use micro-ring modulators to control how much light from each output waveguide is coupled to a pair of detectors. One detector is designated for positive weights and the second detector for negative weights, forming a balanced detector pair. Since the NGRC is designed to apply weights to the optical power in each output waveguide, the optical field coupled from each waveguide should add incoherently at the detector. This can be accomplished by directing $\nout$ spatially separated waveguides to the active detection region of each detector. Alternately, the waveguides could be combined in a multimode waveguide containing at least $\nout$ modes before detection. This design requires $2\nout$ micro-ring modulators. However, since these modulators are only updated at the training stage, they can be slow (e.g., thermal) and the power consumption is amortized across many symbols. Of course, other modulators such as Mach-Zehnder modulators could also be used. Finally, the photocurrent from the two high-speed photodetectors is subtracted in the analog electronic domain to provide the corrected input symbol which can be recorded with a 2-bit digitizer operating at the original symbol rate (e.g., 50 GS/s for the 50 GBd link considered here). 
	
	The proposed PIC-based NGRC shown in \cref{fig:diagram} has several potential advantages compared to DSP-based distortion compensation. First, the NGRC operates on the complex optical field rather than simply the optical power. As we will show, this phase information is crucial to distortion compensation at longer distance. We also find that the NGRC is able to take advantage of this phase information without an LO or co-propagating carrier. Second, by adjusting the spiral delay lines, the NGRC can over-sample the data without requiring high bandwidth electronics. Finally, the output feature vector formation based on spiral delay lines and a mixing region is entirely passive, minimizing power consumption. 
	
	\section{Results: Digital vs. Photonic Distortion Compensation at 10 km}
	
	Our initial simulations were designed to evaluate whether the photonic NGRC feature vector contains the nonlinear transformations and memory required to correct for distortion in an IM/DD link. With this goal in mind, the simulations in this section were performed (1) without the presence of noise, and (2) with full-rank output feature vectors, i.e. $\nout= \nin^2$ (see the discussion around \cref{eq:voutnl}), representing the best-case performance for photonic NGRC. These results provide a baseline for the simulations in Sections \ref{sec:opt1} and \ref{sec:opt2}, where we study the impact of reducing the output feature vector size and including the effect of noise and insertion loss on the equalization performance. 
	
	We first consider a 50 GBd PAM4 link with 10 km propagation length. This length was chosen to exceed the normal operating length of a PAM4 link to investigate the ability of the NGRC to outperform a DSP-based solution. As seen in \cref{fig:diagram}a, the optical power after 10 km is severely distorted; the uncorrected bit error rate was found in simulation to be $\ber = 2 \times 10 ^{-1}$. We attempted to improve this BER using four different equalization schemes: a DSP-based FIR filter, a DSP-based third-order NGRC, the photonic NGRC without an LO, and the photonic NGRC with an LO. The DSP-based FIR filter provides baseline performance (most DSP algorithms begin with a FIR filter), while the DSP-based NGRC is equivalent to a state-of-the-art Volterra based DSP equalizer \cite{diamantopoulosComplexityReductionSecondOrder2019,yuLowcomplexityNonlinearEqualizer2020,zhangLowComplexityVolterra2022,le400GbCWDM42025} (see Appendix \ref{sec:ngrc_model} for more details on the DSP-based correction schemes). In this simulation we used $5 \times 10^3$ training points and $50 \times 10^3$ test points. 
	
	\begin{figure}[h]
		\centering
		\includegraphics{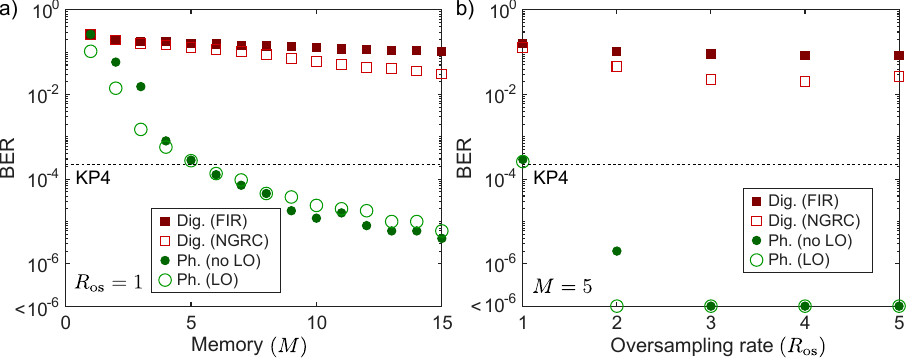}
		\caption{a) Comparison of performance for digital and photonic equalization methods for a 10 km link with no oversampling and a memory spanning from $M = 1$ to 15. b) Comparison of digital and photonic equalization of a 10 km link with a fixed memory $M = 5$ and an oversampling rate ranging from $\ros = 1$ to 5.}\label{fig:ber_vs_memory}
	\end{figure}
	
	\Cref{fig:ber_vs_memory}a shows the BER for these four implementations as a function of memory length with $\ros =1$ (that is, the data is sampled at the encoding data rate of 50 GS/s and $\nin=M$). Each data point represents the mean bit error rate (BER) calculated across 10 independent trials of the distortion and equalization simulation. We found that the photonic NGRCs significantly outperform the DSP-based approaches. In particular, the improved performance of the analog photonic NGRCs compared to the digital NGRC, despite the latter including nonlinearities up to third order, indicates that the inclusion of phase information is more important to equalization performance than additional nonlinearity. The BER for the photonic NGRCs drops below the KP4 threshold of $2.2\times 10^{-4}$ \cite{patraUpdatesConcatenatedFEC2022} when $M \geq 6$, where performance then begins to saturate since dispersion in a 10 km 50 GBd link primarily mixes symbols from $\pm 3$ symbols away. While the presence of an LO improves equalization for $M < 5$, we found in these trials that the LO leads to marginally worse equalization for $M \geq 6$.
	
	As a means of increasing the amount of information in the feature vector while avoiding the inclusion of irrelevant data, we continued to investigate the BER as a function of the oversampling rate, $\ros$, with memory fixed at $M=5$. As shown in \cref{fig:ber_vs_memory}b, both photonic NGRC architectures achieve a BER less than $1\times 10^{-6}$ when the oversampling rate is greater than or equal to $\ros=2$ (LO) or $3$ (no LO). This bound corresponds to the existence of zero errors across all 10 trials of $50\times 10^3$ symbols, with each symbol encoding two bits. Comparing \cref{fig:ber_vs_memory}b and \cref{fig:ber_vs_memory}a, we see that oversampling can improve performance while the input feature vector size, $\nin = M\ros$, is held fixed. In contrast, the DSP-based approaches fail to reach the FEC threshold regardless of the oversampling rate. 
	
	The results in \cref{fig:ber_vs_memory} show that both phase information and oversampling are required for the photonic NGRC to equalize the link distortion. Given that the phase is inaccessible and oversampling is significantly more difficult in DSP-bases IM/DD links, these results mark a clear advantage of the photonic NGRC approach. Finally, while including an LO minimizes errors at a lower oversampling rate compared to the LO-free design, it is much less costly in our photonic NGRC to increase the oversampling rate compared to including an LO. For this reason, the remaining simulations will focus only on LO-free implementations. 
	
	\section{Results: Extending the link range using a photonic NGRC}
	
	In the previous section, we showed that a photonic NGRC without an LO can significantly reduce bit errors in a 10 km 50 GBd noise-free channel. This section demonstrates how a photonic NGRC with $\ros =5$ can be optimized to extend the maximum reach of a link. To show this, we repeated equalization simulations while increasing the link range by factors of two from $L=100$ m to $L=102.4$ km. We expect that longer links, which experience greater pulse spreading due to chromatic dispersion, will benefit from the inclusion of more memory elements. To investigate the role of memory in the NGRC performance, we performed simulations for memory $M=3,5$ and 11. Again, we ignored noise and assumed full-rank ($\nout = \nin^2$) photonic NGRC feature vectors. The BER for these simulated photonic NGRCs is compared against the results for a DSP-based FIR filter or digital NGRC, each with a fixed memory of $M=5$ and oversampling rate of $\ros =5$. In this simulation, we trained on $25 \times 10^3$ points and tested on $50 \times 10^3$ points. This enlarged training set is necessary because output feature vector sizes grow quadratically with memory for photonic NGRCs (e.g. $\nout=3025$ when $M=11$ and $\ros = 5$) and cubically (e.g. $\nout = 3276$ for $M = 5$ and $\ros = 5$) for third-order digital NGRCs, and it is important to ensure there is sufficient training data to optimize weights for all feature vector elements.
	
	\begin{figure}[h]
		\centering
		\includegraphics{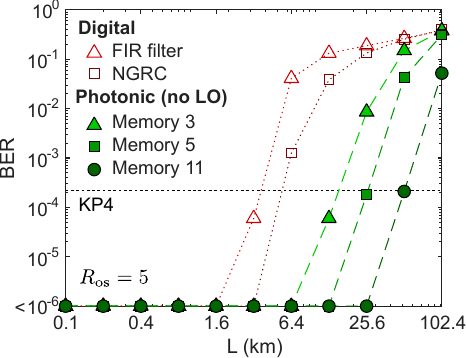}
		\caption{Bit error rate vs. link length for digital methods (FIR filter, polynomial NGRC) compared to a simulated photonic NGRC with memory length varying from $3$ to $11$ data points. The KP4 error correction threshold is shown for comparison.}\label{fig:ber_vs_length}
	\end{figure}
	
	The BER as a function of link length, $L$, is shown in \cref{fig:ber_vs_length}. The simple digital FIR filter is limited to 1.6 km before the BER exceeds the KP4 FEC threshold. The digital NGRC shows modest improvement compared to the FIR filter, enabling operation out to 3.2 km. There is a marked improvement when equalizing with photonic NGRC with only three memory elements, extending the link range to 12.8 km. As we increase memory, the maximum link length increases: the error rate for $M = 5$ remains below the KP4 FEC threshold for a 25.6 km link ($\text{BER} = 1.8 \times 10^{-4}$), and the error rate for $M = 11$ remains below the threshold up to $L = 51.2$ km ($\ber = 2.1\times 10^{-4}$). This result shows the potential for photonic NGRCs to dramatically increase the operating range of IM/DD links. \Cref{fig:ber_vs_length} also highlights an advantage of NGRCs compared to standard RCs: in the NGRC architecture, the memory can be clearly and explicitly controlled to match the memory required by a given task (set here by the chromatic dispersion-induced pulse spreading experienced in the optical fiber). 
	
	\section{Results: Optimizing a photonic NGRC for real-world implementation}\label{sec:opt1}
	
	The results presented so far show that a photonic NGRC has the potential to enable much longer link range than DSP-based solutions due to its access to the optical phase and ability to oversample the transmitted data. However, we saw that the full-rank output feature vector becomes quite large as memory increases. At $M=11$ and $\ros=5$, the ideal output vector had $\nout=3025$ elements. A PIC-based NGRC implementing this design would require 3025 output waveguides and 6050 microring modulators. While this scale of integration is not impossible in principle, reducing $\nout$ would help to simplify the PIC design and reduce cost. In this section, we detail the effects of limiting $\nout$ by evaluating the BER for a 10 km link equalized by an LO-free photonic NGRC with $M = 5$ and $\ros = 5$ as $\nout$ is reduced from its full-rank value. After, we demonstrate how a photonic NGRC may be designed under a fixed practical constraint on $\nout$.
	
\begin{figure}[h]
	\centering
	\includegraphics{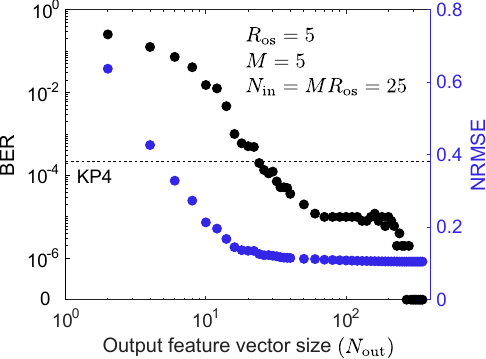}
	\caption{BER (left axis) and NRMSE (right axis) versus output feature vector dimension $\nout$ for a memory $M = 5$ and oversampling rate $\ros = 5$.}\label{fig:ber_vs_M}
	\end{figure}

	The BER for the photonic NGRC without an LO is shown in \cref{fig:ber_vs_M} as a function of $\nout$. \Cref{fig:ber_vs_M} also shows the normalized root-mean square error (NRMSE) between the recovered signal and the input signal level, which provides a complementary view of the distortion compensation process. As $\nout$ increases from 2 to $\sim$ 10, we see a sharp decrease in the NRMSE. However, this leads to a modest decrease in the BER which lags the NRMSE. The BER reaches an initial plateau for $\nout =50$ before dropping to an error of zero for $\nout \gtrsim 200$, while the NRMSE shows only modest improvements beyond $\nout \approx 20$. Most importantly, we find that the FEC threshold is crossed at $\nout = 30$. 
	
\begin{figure}[h]
		\centering
		\includegraphics{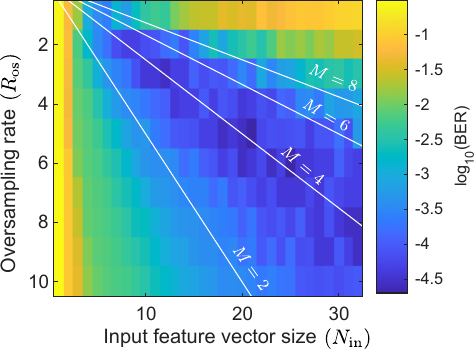}
		\caption{BER on a logarithmic scale as a function of oversampling rate $\ros$ and input feature vector size $\nin$ for a fixed output dimension $\nout = 32$. White lines indicate the parameters with integer values of memory $M$.}\label{fig:ber_2d_fixed_dimension}
	\end{figure}

	\Cref{fig:ber_vs_M} showed that is is possible to reduce $\nout$ for a fixed memory and oversampling rate and still achieve satisfactory equalization performance. In a practical design, however, it is more natural to first impose a constraint on the output feature vector size, $\nout$, then optimize the memory and oversampling simultaneously. We performed this procedure on the same 10 km link by fixing $\nout=32$ and varying these hyperparameters to determine which combination yielded the lowest error rate. \Cref{fig:ber_2d_fixed_dimension} shows the resulting BER as a function of $\ros$ and $\nin$. We vary $\nin$ instead of $M$ because this scheme allows us to include input feature vectors with ``fractional'' memory instead of restricting our memory increment size to the symbol rate. In each case, the memory can be calculated as $M=\nin/\ros$  and the white contour lines track the BER at fixed memory of 2, 4, 6, or 8. 
	
	We observe the fewest errors when $M \approx 4$, which captures most of the intersymbol interference induced by the DAC response and chromatic dispersion in the 10 km link. Below $M=4$, the feature vector contains insufficient information to reconstruct the undistorted signal. We also find degraded performance with too much memory: for $M>6$, the feature vector contains samples that are uncorrelated with the data point to be reconstructed, causing an increase in the BER. This trend is consistent with the behavior in \cref{fig:ber_vs_memory}a, compounded by an increase in the difficulty of extracting relevant information from rank-deficient feature vectors using linear regression. We found the optimal feature vector to have $M=4.2$ and $\ros=5$, which produced a BER of $2\times 10^{-5}$, well below the FEC threshold. This procedure shows how a realistic PIC-based photonic NGRC can be optimized for a target link length.

	\section{Results: Optimizing a photonic NGRC in the presence of noise}\label{sec:opt2}
	
	In the previous section, we optimized an NGRC under the constraint that $\nout$ is limited to 32 (corresponding to 32 output waveguides and 64 modulators, which is well within the integration capabilities of photonic foundries \cite{daudlinThreedimensionalPhotonicIntegration2025}). We found that for a 10 km link, the optimal design uses $\nin=21$ input delays with $M=4.2$ and $\ros =5$. Here, we simulate how the optimized photonic NGRC operates in the presence of shot noise and ASE introduced by the EDFA shown in \cref{fig:diagram}a. 
	
	The simulation was performed as follows: We set the initial laser power to 2 mW and modeled the data encoding process as described in Appendix \ref{sec:dis}. After propagation through a 10 km fiber (with loss of 0.2 dB/km), the signal was amplified by an Erbium-doped fiber amplifier (EDFA) to compensate for fiber channel and PIC insertion losses. This amplification introduces ASE to the optical signal with a noise factor $\text{NF}=5$ (see Section \ref{sec:ASE} in Appendix \ref{sec:dis}). We studied the BER as a function of EDFA gain, ranging from 0 dB (i.e., without an EDFA) to 30 dB. The amplified signal was then coupled onto the PIC. We assumed a total insertion loss of 4 dB, including 2 dB coupling loss and 2 dB waveguide propagation loss. Note that the longest optical path light can travel on the PIC, including the longest delay line, the mixing region, and the output coupling, is $\sim$ 5 mm. Considering standard waveguide loss of $\sim$ 0.5 dB/cm \cite{suSiliconPhotonicPlatform2020}, this corresponds to just 0.25 dB of loss. Thus, our 2 dB estimate provides some margin for fabrication imperfections. Finally, we model a finite-bandwidth photodetection process including shot noise (see Section \ref{sec:shot} in Appendix \ref{sec:dis} for details). We repeated the simulation for 10 trials at each EDFA gain level.  
	
		\begin{figure}[h]
		\centering
		\includegraphics{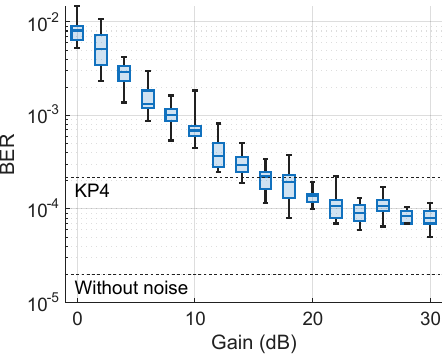}
		\caption{BER versus gain for a simulated PIC NGRC with ASE and shot noise. Equalization is applied to a 10 km link and the NGRC parameters are fixed to have output dimension $\nout = 32$ with $M = 4.2$ and $\ros = 5$.}\label{fig:ser_vs_gain}
		\end{figure}

	The BER is shown as a function of EDFA gain in \cref{fig:ser_vs_gain}. The blue lines denote the mean, 25\%, and 75\% thresholds while the horizontal black lines mark the maximum and minimum BER found at each gain value. We found that at 0 dB gain (i.e., without the EDFA), shot noise increases the BER from $2 \times 10^{-5}$ (as found without noise in \cref{fig:ber_2d_fixed_dimension}) to $8\times 10^{-3}$, which is above the FEC threshold for KP4. The BER decreases as the EDFA gain increases, falling below the FEC threshold at a gain of $\sim$20 dB or higher. The satisfaction of the threshold is true for every trial in this set except for one point at a gain of 22 dB. As gain increases further, the system becomes ASE noise dominated and we find no significant further improvement. This result shows that, with the application of reasonable gain, the insertion loss on the PIC based photonic NGRC can be overcome and communication with BER below the FEC threshold is possible.
	
	\section{Conclusion}

	In this work, we proposed a nonlinear pluggable optic (NLPO) based on an analog photonic NGRC. Our simulations showed that the NGRC-based NLPO enables longer range operation than DSP based IM/DD links while maintaining the low power consumption and low latency of LPO transceivers. Compared to existing LPO transceivers, the NLPO supports orders of magnitude longer operating range, enables use in the C-band with non-zero dispersion, and does not require specialized electronics to linearize the encoded data.
	
	This work also identified two major advantages of photonic NGRCs. First, the NGRC is able to passively perform calculations on the optical phase without requiring an LO or co-propagating carrier. Second, the NGRC is able to oversample the data well beyond the symbol rate without requiring prohibitively expensive electronics. We found that oversampling is critical to enabling error-free communication. After demonstrating that an ideal photonic NGRC (using a full-rank feature vector in the absence of noise) can compensate for distortion in an IM/DD link, we also showed how the NGRC can be optimized for a realistic implementation by pruning the feature vector and incorporating an amplifier to compensate for insertion loss. We are currently working toward an experimental realization of a photonic NGRC based NLPO and hope to present that in the future.
	
	\acknowledgments
	The authors acknowledge support from the U.S. Naval Research Laboratory (6.2 Base Funding). We thank George Valley for useful discussions on next-generation reservoir computing.
	\appendix
	\section{Simulation of the optical link}
	\label{sec:dis}
	This appendix details the simulations used to model a 50 GBd PAM4 optical communication link. The procedure involves generating a sequence of random data points sampled from a four-level distribution, encoding this information onto an optical carrier, and simulating the optical propagation and subsequent photodetection at the receiver end. In particular, we simulate the effects of channel distortion due to: (1) the RF-domain DAC response, (2) optical modulation and propagation, and (3) noise.
	
		\subsection{DAC response}
	\label{sec:DAC}

	We begin the simulation by generating data to be transmitted, selecting a quantity $N_{\text{data}}$ samples from a uniform random distribution with elements $S = \{-1, -1/3, 1/3, 1\}$, corresponding to the (ordered) Gray-coded bit strings $\{00,01,10,11\}$. This collection of undistorted data samples, denoted by $u[k]$ at time step $k$, is represented by the sample-and-hold waveform
	\begin{equation}\label{eq:ut}
		u(t) = V_{0}\sum_{i = 1}^{N_{\text{data}}} u[k] \text{rect}\left(\frac{t - kT_0}{T_0}\right),
	\end{equation}
	with symbol period $T_0 = 20$ ps. The scaling factor $V_0$ sets the maximum value of the idealized voltage waveform (i.e., $\pm V_0$ are the voltages that encode symbols $\pm 1$). This signal is converted into a finite-bandwidth analog voltage using a digital-to-analog converter (DAC). Ignoring DAC nonlinearitites, we take the output voltage to be characterized by the convolution $V(t) = u(t)*h(t)$ for the DAC impulse response $h(t)$. In simulation, we apply the discrete-time convolution by defining the DAC output voltage
	\begin{equation}\label{eq:Vin}
		V_{\text{in}}(t) = V_0\sum_{k = 1}^{N_{\text{data}}} u[k]  h_s(t - kT_0),
	\end{equation} 
	where $h_s = h(t)*\text{rect}(t/T_0)$ is the DAC response to a single symbol of width $T_0$.

	We experimentally generated a realistic symbol response $h_s(t)$ in the following way. An arbitrary waveform generator with 15 GHz 3 dB RF bandwidth was programmed to create a square pulse of duration $2 T_0$, comprising two samples at a 50 GS/s sampling rate, and the resulting waveform was measured on an oscilloscope at a rate of 100 GS/s. The time axis of the trace was re-scaled by a factor of two so that the resulting waveform simulates the input of a square pulse of width $T_0$ with an effective RF bandwidth of 30 GHz. The response $h_s$ was then interpolated at the simulation sampling rate and the resulting voltage was computed as in \cref{eq:Vin}.
	
\begin{figure}[h]
		\centering
		\includegraphics{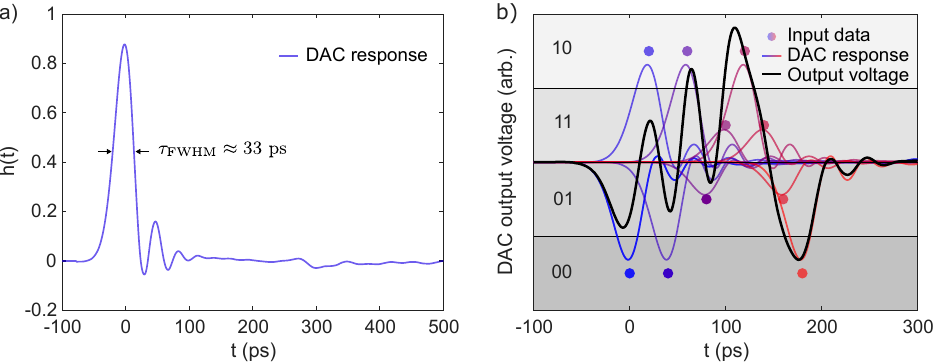}
		\caption{a) DAC impulse response used for simulations, found by rescaling the time axis of the oscilloscope trace measuring the output of an arbitrary waveform generator with 15 GHz RF bandwidth. b) A plot showing the effects of the DAC impulse response. Horizontal lines indicate thresholds for assigning the sampled output voltage to a Gray-coded two-bit sequence.}\label{fig:dac_response}
	\end{figure}

	The simulated response is shown in \cref{fig:dac_response}a. The response has a width of $\approx 33$ ps (FWHM), which is longer than the symbol period $T_0 = 20 $ ps. Additional ringing occurs beyond 100 ps, with a long-tailed distortion that continues for $> 500$ ps. \Cref{fig:dac_response}b shows how the DAC response distorts a series of random data points, introducing significant intersymbol interference (ISI). This interference introduces errors in the electrical domain, even before optical distortion. We can see these errors by observing that each shaded region represents the range for which the output voltage, sampled at the same time as each input data point, is assigned to a given 2-bit sequence. For the first three data points in \cref{fig:dac_response}b, the sampled value of the output voltage lies across a single threshold from the true data point, indicating a single bit error in the Gray-coded scheme for each of these three points.
	
	We did not isolate the distorting effects of the oscilloscope circuitry when generating $h_s(t)$, meaning it is assumed that the measured output is representative of the DAC impulse response. It is also possible that a voltage generator with higher bandwidth and more sophisticated filtering will induce fewer errors in the electrical domain than what we see in \cref{fig:dac_response}b. However, despite the additional distortion due to these limitations, we find in this work that photonic NGRC is still capable of reducing bit error rates to below FEC thresholds.
	
	\subsection{Encoding and propagation}
	The data encoding voltage, $V_{\text{in}}(t)$, defined in \cref{eq:Vin}, is converted to a simulated optical waveform by applying the electrical-to-optical response function for an MZI-based EOM biased at quadrature:
	\begin{equation}\label{eq:EOM}
		\ta(t) = A_0 \cos\left(\frac{V_{\text{in}}(t)}{2 V_\pi} - \frac{\pi}{4}\right) \leftrightarrow |\ta(t)|^2 = \frac{|A_0|^2}{2}\left(1 + \sin\left(\frac{V_\text{in}(t)}{V_\pi}\right)\right),
	\end{equation}
	where $V_\pi$ is the modulator half-wave voltage. The output $\tilde{A}(t)$ is the complex-valued electric field normalized so that $|\tilde{A}(t)|^2$ is the instantaneous optical power in the fiber mode. The initial power of the laser source is specified by $|A_0|^2$.
	
	The right-hand side of \Cref{eq:EOM} indicates that the voltage relates to the resulting optical power by a nonlinear relationship, which further distorts the signal beyond the DAC-induced ISI. We set the voltage amplitude to a value $V_0 = V_\pi/4$, which provides a reasonable trade-off between nonlinearity of the modulation response and SNR in a noisy simulation. Note that \cref{eq:EOM} is an idealized response for a single modulator type; real-world modulation may also introduce linear filtering and phase chirp. It has been shown that a Volterra nonlinear equalizer (equivalent to digital NGRC) is capable of compensating for some degree of chirp \cite{le400GbCWDM42025}. While outside the scope of this work, we believe that the phase information retained in the photonic NGRC presented here makes it especially capable of correcting for modulator chirp.
	
	The propagation of the input signal $\tilde{A}(t)$ through the link fiber is simulated by the nonlinear Schr{\"o}dinger equation
	\begin{equation}\label{eq:schr}
		\frac{\partial \ta(z,t)}{\partial z} = i\beta_2 \frac{\partial^2 \ta(z,t)}{\partial t^2} + i\gamma|\ta(z,t)|^2 \ta(z,t) -\frac{\sigma}{2}\ta(z,t),
	\end{equation}
	integrated by the split-step Fourier over a link length $L$ to compute the received field 
	\begin{equation}\label{eq:Adist}
		\ta_{\text{dist}}(t) = \ta(L,t).
	\end{equation}
	Simulating propagation according to \Cref{eq:schr} introduces distortions via group velocity dispersion (with $\beta = -21.4 \text{\ ps}^2/\text{km}$) and the instantaneous Kerr effect (with $\gamma = 1.2  \text{\ W}^{-1} \text{\ km}^{-1}$). In the case that noise is considered, the loss incurred with $\sigma = 4.6 \times 10^{-2} \text{\ km}^{-1}$ leads to additional communication errors by degrading the SNR. 
	
	Finally, we denote the field reaching the PIC as $\tilde{A}_{\text{rx}}(t)$. This choice allows us to generalize the analysis for noise-free and noisy simulation. Without noise, we simply have that $\ta_{\text{rx}} = \ta_{\text{dist}}$. We provide a formula for defining $\ta_{\text{rx}}$ in the presence of ASE noise in Section \ref{sec:ASE}.
	\subsection{Discretization and thresholding}
	Throughout the remaining sections of Appendix \ref{sec:dis} and \ref{sec:ngrc_model}, we present simulation details performed on the discrete-time grid with spacing $\Delta t_s$. This simulation time step is related to the symbol period $T_0$ by the relationship $T_0 = N_{\text{sps}} \Delta t_s$, where $N_{\text{sps}}$ is the integer-valued number of simulation samples per symbol period. In this work, we use $N_{\text{sps}} \approx 20$, which must vary as we change the oversampling rate to ensure that $\ros$ is a divisor of $N_{\text{sps}}$ for consistent sampling of the PAM4 data.
	
	In continuous time, the recorded signal at the receiving end of the link is determined by the voltage output, $V_{\text{out}}(t)$, of a photodetector. For an uncorrected link, this quantity is proportional to the received optical power that has been distorted by the channel. In the case of ideal detection, this voltage waveform is computed from the received optical signal by $V_{\text{out}}(t) = |\ta_{\text{rx}}(t)|^2$ (the case with realistic detection is detailed in Section \ref{sec:noise} of this appendix). The discretized form of this output is defined by
	\begin{equation}\label{eq:d_sample}
		V_{\text{out}}[l] := V_{\text{out}}(l\Delta t_s).
	\end{equation}
	To extract the received distorted symbols, denoted by $d[k]$, from \cref{eq:d_sample}, we subsample at the symbol time step by
	\begin{equation}\label{eq:dk}
		d[k] = V_{\text{out}}[l_0 + kN_{\text{sps}}].
	\end{equation}
	To avoid ambiguity, we reserve the index $k$ to be applied only to the symbol indices $k = 1, \dots, N_{\text{data}}$, and any discrete function $f[k] := f(kT_0)$ is spaced by the symbol period. \Cref{eq:dk} applies a sampling offset $l_0$ which selects the time when the sampling of data points begins. In practice, we vary the value of $l_0$ until the lowest BER is achieved for both uncorrected and NGRC-equalized links.
	
	The goal of this work is to apply optical transformation to the complex field $\ta_{\text{rx}}$ so that the final measured voltage $V_{\text{out}}[l]$ provides a more faithful reconstruction of the undistorted data. For the input data signal defined by $u[k]$, we define our reconstruction as the quantity $\hat{u}[k]$, sampled the same as in \cref{eq:dk}. As a final inference step, this reconstructed sequence is thresholded according to the function
	\begin{equation}\label{eq:dn}
		\overline{u}[k] = f_T \left(\hat{u}[k]\right)
	\end{equation}
	to map the sampled data back to one of the four PAM4 symbol levels. For NGRC operation, we optimize $f_T$ according to a nonlinear optimization scheme to find the thresholds that yield the lowest BER.

	\subsection{Noise and detection models}\label{sec:noise}
	In this work, we vary the gain of a simulated EDFA to optimize the performance of a photonic NGRC in the presence of noise. In particular, we include the effects of ASE noise due to the amplification and shot noise at the detection stage.
	\subsubsection{ASE noise}\label{sec:ASE}

	ASE noise is simulated in the optical domain as an additive distortion applied to $\ta_{\text{dist}}[l]$, which is the the distorted optical signal at the receive side at time step $l$, sampled at the simulation time step $\Delta t_s$.  Pre-amplifying this received signal by a factor $G$ introduces noise with power spectral density
	\begin{equation}\label{eq:sase}
		S_{\text{ASE}} = \frac{n_{\text{sp}}}{2}(G - 1)h\nu,
	\end{equation}
	where $n_{\text{sp}}$ is the spontaneous emission factor defined by $n_{\text{sp}} = 10^{\text{NF}/10}$ and $\text{NF}$ is the noise factor. In this work, we fix $\text{NF} = 5$. We define the optical noise field, $\ta_{\text{ASE}}$, with the power spectral density of \cref{eq:sase} by
		\begin{equation}
		\ta_{\text{ASE}}[l] = \sqrt{\frac{S_{\text{ASE}}}{\Delta t_s}} Z[l],
	\end{equation}
	where $Z[l]$ is a complex random field with in-phase and quadrature components
	\begin{equation}
		Z[l] = \frac{X[l] + iY[l]}{\sqrt{2}} 
	\end{equation}
	sampled from Gaussian random variables $X[l]$ and $Y[l]$ with zero mean and unit variance. This noise field is added to the signal to yield the total field
	\begin{equation}
		\ta_{\text{tot}}[l] = \ta_{\text{ASE}}[l] + \ta_{\text{dist}}[l].
	\end{equation}
	To improve SNR, we apply a simulated bandpass filter to find the received field 
	\begin{equation}\label{eq:afilt}
		\ta_{\text{rx}}[l] = \ta_{\text{tot}}[l]*h_{\text{BPF}}[l],
	\end{equation}
	where $h_{\text{BPF}}[l]$ is the inverse Fourier transform with super-Gaussian frequency response
	\begin{equation}
		H_{\text{BPF}}[f] = \exp\left[-\left(\frac{2 f}{B}\right)^{2m}\right],
	\end{equation}
    with order $m = 4$. We choose the two-sided bandwidth to be $B = 100$ GHz to reduce noise power without introducing significant additional distortions to the input signal.
	\subsubsection{Photodetection and shot noise}\label{sec:shot}
	Shot noise is accounted for at the detection stage of the simulation; instead of the idealized square-law response $V_{\text{out}}(t) = |\ta|^2$ for an optical field $\ta$, we model detector output current by the random arrival of photoexcited carriers. The photocurrent at simulation time step $l$ is
	\begin{equation}\label{eq:Idet}
		I[l] = \frac{q C[l]}{\Delta t_s},
	\end{equation}
	where $q$ is the charge of an electron and $C[l]$ is the number of photodetector counts during the simulation interval of duration $\Delta t_s$. The counts are sampled from the Poisson distribution characterized by the mean value
	\begin{equation}\label{eq:shot}
		C[l] \sim \text{Poisson}\left(\frac{RP_{\text{opt}}[l]  \Delta t_s}{q}\right),
	\end{equation}
	proportional to the optical power $P_{\text{opt}}[l] = |\ta[l]|^2$ of the signal, and the detector responsitivity, $R$, which we fix to be $R = 1$ A/W. For an uncorrected link, the detected signal is the pre-amplified power $P_{\text{opt}} = |\ta_{\text{rx}}[l]|^2$ defined in \cref{eq:afilt}. For NGRC operation, the detected power at the positive and negative detectors in the balanced detection is given by
	\begin{equation}
		P_{\text{opt},\pm}[l] = \sum_{n = 1}^{\nout}w_{\pm,n}\left|A_{\text{mix},n}[l]\right|^2,
	\end{equation}
	where $w_{\pm,n}$ are the positive and negative weights, respectively, applied to the feature vectors generated according to \cref{eq:Amix}. As mentioned in the main text, we assume that the waveguide traces for each path to the detector are spatially separated so that the power depends on the incoherent weighted sum of the individual powers $\left|A_{\text{mix},n}[l]\right|^2 = \voutn[l]$, which are the output feature vectors at the simulation time step $l$. In the case of both single-ended and balanced detection, the combined noise model accounts for shot noise arising from both the communication data and the ASE noise present in the received optical signal.
	
	The current defined by \cref{eq:Idet,eq:shot} is converted to a voltage $V_{\text{out}}[l]$ by a simulated transimpedance amplifier with impulse response
	\begin{equation}\label{eq:Vout_det}
		V_\text{out}[l] = h_{\text{TIA}}[l]*I[l].
	\end{equation}
	The finite-bandwidth response function $h_{\text{TIA}}[l]$ is defined as the discrete-time inverse Fourier transform of the two-pole filter
	\begin{equation}
		H_{\text{TIA}}(\omega) = \frac{G_0 \omega_0^2}{\omega^2 +  2\zeta \omega \omega_0 + \omega_0^2}.
	\end{equation}
	Defining $\omega_0 = 2\pi f_c$, we chose the cutoff frequency $f_c = 30$ GHz and a damping parameter $\zeta = 0.3$ to achieve a realistic 37 GHz FWHM bandwidth and ringing comparable to that seen in real detectors. The bandwidth was chosen to be greater than that of the encoding DAC to ensure noise filtering without introducing a significant amount of additional ISI. Ignoring RF amplifier noise, the transimpedance gain $G_0$ has no impact on the SNR so we fix $G_0 = 1\ \Omega$.

	\section{Mathematical model of the photonic NGRC}\label{sec:ngrc_model}

	Appendix \ref{sec:dis} modelled the optical transmission of PAM4 data across a fiber link, demonstrating the distortion due to digital-to-analog conversion, modulator nonlinearity, and optical propagation. In the case of realistic simulated detection, we showed in Section \ref{sec:noise} that the signal is further degraded by the detector impulse response and noise. This appendix provides details about the NGRC-based equalization procedure used to compensate for these distortions. We begin with a brief overview of the equalization procedure, then provide the theoretical and simulation details for digital NGRC and photonic NGRC introduced in Section \ref{sec:pic}. Finally, we present an example of the photonic NGRC mixing structure that generates output feature vectors for PIC-based equalization.
	
	\subsection{Equalization architecture and training procedure}
	
	In this work, we compare the DSP-based methods of FIR filtering and digital NGRC to the photonic NGRC with and without a local oscillator contribution. In each case, we begin the operation by creating an input feature vector $\mathbf{v}_{\text{in}}[l]$ defined on the time step $l$ sampled at the simulation interval $\Delta t_s$. In this section, we elaborate on the distinction between the input feature vectors used in digital and photonic approaches; namely, that the former derives elements from the optical power and the latter derives its elements from the complex electric field. In each case, the input feature vector contains a quantity $\nin$ delay samples separated by a period $T_0/\ros$, representing an oversampling factor of $\ros$ with respect to the input data.
	
	These input feature vectors are used to generate output feature vectors $\mathbf{v}_{\text{out}}[l]$ by a nonlinear transformation. The output feature vector is finally resampled at the original data rate $f_0$ to correspond to symbol $k$ by
	\begin{equation}\label{eq:vsm}
		\mathbf{v}_{\text{out}}[k] := \mathbf{v}_{\text{out}}[l_0 + N_{\text{sps}}k],
	\end{equation}
	where $N_{\text{sps}}$ is the number of samples per symbol introduced in Appendix \ref{sec:dis} and $l_0$ is the sampling offset that determines how the memory structure is centered within the feature vector. To make this offset clear, consider the uncorrected link for which $\mathbf{v}_{\text{out}}[l] = V_{\text{out}}[l]$, representing the distorted optical power (see \cref{eq:d_sample}). Then, supposing that $M = 5$ and $\ros = 1$, the value $l_0 = -3 N_{\text{sps}}$ fixes the memory window to be symmetric about the distorted data point $d[k]$, since then $\voutn[k] = V_{\text{out}}[(n + k - 3)N_{\text{sps}}]$ with $n = 1,\dots,5$. 
	
	In practice, the value of $l_0$ is optimized during the training process, which proceeds in the following way. Generating a data set comprising $N_{\text{train}}$ and $N_{\text{test}}$ random symbols, we simulate the transmission of this data through the link according to Appendix \ref{sec:dis}. From the distorted signal at the receiving end of the link, we generate output feature vectors at each symbol time $k$ according to \cref{eq:vsm} for a given $l_0$. This output feature vector is weighted to find the approximation of the undistorted signal $\hat{u}[k]$,
	\begin{equation}
		\hat{u}[k] = w_0 + \mathbf{w}\cdot \vout[k],
	\end{equation}
	where $\mathbf{w}$ are the weights applied to the feature vector and $w_0$ is a DC offset. The weights and offset are determined by linear regression to reduce the squared error between the true signal $u[k]$ and the reconstructed signal $\hat{u}[k]$ for some fixed set of training points. Once we have $\hat{u}[k]$, the thresholding function $f_T$ is optimized so that the reconstructed symbols $\overline{u}[k] = f_T(\hat{u}[k])$ (see \cref{eq:dn}) exhibit the lowest bit error rate (BER) for the training data. When applying these thresholds, we may ignore the value $w_0$ and account for the offset by shifting the threshold values. This process is repeated for values of $l_0$ which span over the memory duration of the input feature vector and the result with the best performance is used in the test phase. In the end, the number of errors that occur when applying the optimized weights and thresholds during the testing phase determines the BER. The Gray coding scheme ensures that crossing a single threshold results in a single bit error.
	
	\subsection{Digital NGRC}

	Presently, PAM4 channel equalization is usually accomplished in DSP by a simple FIR filter or Volterra equalizer (equivalent to a digital NGRC). Since digital methods can only be applied after converting the optical signal into a voltage by photodetection, the input feature vector must always be a real quantity proportional to the measured optical power. In this section, we first describe the simulated input feature vectors for DSP-based equalization. After, we introduce the output feature vectors used in the FIR filter and third-order digital NGRC approaches.
	
	\subsubsection{Digital input feature vector}
	For both FIR filtering and the application of a digital NGRC, the $n$th component of the input feature vector is determined from the distorted optical signal by
	\begin{equation}
		v_{\text{in},n}^{\text{dig.}}[l] = \begin{cases}
			1 \quad & n = 0 \\
			V_{\text{out}}\left[l + n\frac{N_{\text{sps}}}{\ros}\right] & n = 1,\dots,\nin,
		\end{cases}
	\end{equation}
	where $V_{\text{out}}[l]$ is the simulated output voltage, proportional to the received optical power, given in \cref{eq:Vout_det}. Note that we include a constant element at index $0$ for a total of $\nin + 1$ elements. In this work, digital equalization is only applied with a noise-free ideal detection response for which $V_{\text{out}}[l] = |\ta_{\text{rx}}[l]|^2$. As a result, this vector is simply the optical power sampled at $\nin$ time points separated by the interval $T_0/\ros$. Note that, while we simulate oversampling in the digital NGRC methods for a fair comparison of simulated performance, practical implementation of this oversampling introduces the difficulty of digitization and processing at a multiple $\ros$ of the 50 GBd data rate.
	
	\subsubsection{Digital output feature vectors}
	
	For a FIR filter, the output feature vector is simply the input feature vector
	\begin{equation}
		v_{\text{out},n}^{\text{FIR}} = v_{\text{in},n}^{\text{dig.}}.
	\end{equation}
	We see that the usual application of the weights by $\hat{u}[k] = \mathbf{w}\cdot \vout[l_0 + N_{\text{sps}}k]$, which is simply a weighted sum of adjacent time samples, is simply the equation of a FIR filter (with optional DC offset) applied to the original photovoltage signal.
	
	In contrast to the FIR filter, the digital NGRC applies a nonlinear transformation which includes all multiplicative combinations of elements up to a given order. In this work, we consider a third-order digital NGRC which is characterized by the output feature vector
	\begin{equation}\label{eq:ngrc3}
		\voutn^{\text{NGRC}}[l] = \sum_{i = 0}^{\nin} \sum_{j \geq i}^{\nin}\sum_{k \geq j}^{\nin} v^{\text{dig.}}_{\text{in},i}[l] v^{\text{dig.}}_{\text{in},j}[l] v^{\text{dig.}}_{\text{in},k}[l],
	\end{equation}
	with the sum indices enforcing the condition that each unique combination of three elements appears only once. The presence of the unit element in the input feature vector ensures that we represent all constant, linear, quadratic, and cubic terms in the output feature vector (i.e., linear terms are constructed using two identity elements and quadratic terms are constructed using one identity element). Note that the components $v_{\text{in},i}[l]$ are delay samples with temporal spacing $T_0/\ros$ (equivalent to $N_{\text{sps}}/\ros$ in the discrete-time simulation), so that the feature vector contains all third-order combinations of each delay element contained in the input feature vector. In total, there are $\nout = (\nin + 1)(\nin + 2)(\nin + 3)/6$ elements in the feature vector, equal to the number of ways to select unique unordered combinations of three elements from $\nin$ with replacement.
		
	\subsection{Photonic NGRC}
	As shown in \cref{fig:diagram} in the main text, the distorted signal after propagation through the link, $\ta_{\text{dist}}(t)$, is amplified to form the received field $\tilde{A}_{\text{rx}}(t)$ and coupled into a single-mode waveguide on a PIC. This signal is split into $\nin$ paths with relative delay $T_0/R_{\text{os}}$ between each path. The input traces couple into a symmetric multimode mixing region with $\nout$ input and output waveguides. In this work, we always assume that $\nout \geq \nin$ so that $\nout - \nin$ traces are disconnected on the input side. The number of disconnected traces is reduced by one when an LO is included. The field of each of the waveguides on the input side, including or excluding a local oscillator, serves as the $n$th component of the input feature vector $\vinn[l]$ at simulation time step $l$. Note that the input feature vector is denoted with a tilde to indicate that it is a complex quantity. The components of this vector are
	\begin{align} \label{eq:vinq}
		\vinn[l] = \begin{cases}
			\sqrt{P_{\text{LO}}}\delta_{\text{LO}} &\quad n = 0 \\
			\ta_{\text{rx}}\left[l + n\frac{N_{\text{sps}}}{\ros}\right] &\quad n = 1, \dots, \nin\\
			0 &\quad n = \nin + 1,\dots,\nout,
		\end{cases}
	\end{align} 
	where $\delta_{LO} = 1$ if an LO is included and $0$ otherwise. The quantity $P_{\text{LO}}$ is the LO optical power. This input feature vector contains a memory of duration $M = \nin/\ros$, equal to the number of symbol periods spanned by the feature vector.

   \subsubsection{The output feature vector}
   	The transformation applied to the $\nout$-dimensional input feature vector in \cref{eq:vinq} by propagating through the mixing region is modeled by the matrix multiplication
   \begin{equation}\label{eq:Amix}
   	\bta_{\text{mix}}[l] = \tilde{U} \vin[l] \leftrightarrow \ta_{\text{mix},n}[l] = \sum_{i = 1}^{\nout} \tilde{U}_{ni}\tilde{v}_{\text{in},i}[l],
   \end{equation}
   where $\tilde{U}$ is a random $\nout\times \nout$ unitary matrix. Each component $\ta_{\text{mix},n}[k]$ represents the normalized complex electric field profile of output waveguide $n$ at simulation time step $l$. 
   
   To form the output feature vector, each of these outputs is measured via square-law photodetection. In the case of idealized detection with no filtering, the components of the output feature vector are
	\begin{align}\label{eq:voutnl}
		\voutn[l]	= \sum_{i,j = 1}^{\nin} \left(\tilde{U}^*_{ni}\tilde{U}_{nj}\right)\tilde{v}^*_{\text{in},i}[l] \tilde{v}_{\text{in},j}[l], \quad n = 1,\dots,\nout.
	\end{align}
	As we increase $\nout$ by introducing additional mixing waveguides, we eventually reach a value for which any additional output $n > D$ is linearly dependent on the elements $\voutn[l]$ for $n \leq D$. This quantity $D$ is the dimension of the vector space of which the output feature vector is a member. The number of linearly independent dimensions is made apparent if we separate the input feature vectors into their real and imaginary parts, i.e. $\tilde{v}_{\text{in},i} = v_{\text{in},i}^R + i v_{\text{in},i}^I$. In the case with no local oscillator, applying this expansion to \cref{eq:voutnl} and separating diagonal $(i = j)$ and cross $(i \neq j)$ terms yields three sums:
	\begin{align}\label{eq:voutnl}
		v_{\text{out},n}^{\text{no LO}}[l]&= \sum_{i =1}^{\nin} |\tilde{U}_{ni}|^2 |\tilde{v}_{\text{in},i}[l]|^2 \nonumber\\
		&+\sum_{i,j > i}^{\nin} 2\text{Re}\left\{\tilde{U}^*_{ni}\tilde{U}_{nj}\right\}\Big(v^R_{\text{in},i}[l]v^R_{\text{in},j}[l] + v^I_{\text{in},i}[l]v^I_{\text{in},j}[l]\Big) \nonumber\\
		&+ \sum_{i,j > i}^{\nin} 2\text{Im}\left\{\tilde{U}^*_{ni}\tilde{U}_{nj}\right\}\Big(v^R_{\text{in},i}[l]v^I_{\text{in},j}[l] -v^I_{\text{in},i}[l]v^R_{\text{in},j}[l]\Big)
	\end{align}
	for $n = 1,\dots,\nout$. We deduce the dimension $D$ of the output vector space by interpreting the parenthetical terms in \Cref{eq:voutnl} as its basis vectors. In the order shown in \cref{eq:voutnl}, the three sums contain $\nin$, $\nin(\nin-1)/2$, and $\nin(\nin-1)/2$ unique terms, for a total of $\nin^2$ basis vectors. Consequently, each feature vector element $v_{\text{out},n}[l]$ for a given $n$ represents an $\nin^2$-dimensional vector in the space. Assuming linear independence between the vectors $v_{\text{out},n}$ is imposed by the expansion coefficients (computed from elements of $\tilde{U}$), we find that increasing $\nout > \nin^2$ necessarily generates a vector that is linearly dependent on the previous $\nin^2$ vectors. For this reason, we say that a feature vector is full-rank when $\nout = D = \nin^2$ (with no LO). The consequence of being full-rank is that any basis element of the vector space may be isolated using an appropriate linear combination of output vectors. Thus, once $\nout \geq D$, we may (for example) determine a set of weights to apply to the output vectors such that $\mathbf{w}\cdot\mathbf{v}_{\text{out},n}[l] = |\tilde{v}_{{in},i}[l]|^2$ for some index $i$.
	
	NGRCs with an LO gain additional terms in the output feature vector, given by
	\begin{equation}\label{eq:voutlo}
		v^{\text{with LO}}_{\text{out},n}[l] = v_\text{out,n}^{\text{no LO}}[l] + |U_{n0}|^2 + \sum_{i = 1}^{\nin} 2\text{Re}\left\{\tilde{U}_{ni}^*U_{n0}\right\}v_{\text{in},i}^R[l] - \sum_{i = 1}^{\nin} 2\text{Im}\left\{\tilde{U}_{ni}^*U_{n0}\right\}v_{\text{in},i}^I[l].
	\end{equation}
	According to \cref{eq:voutlo}, the LO introduces a single constant term along with $2\nin$ additional basis vectors equal to the real and imaginary parts, respectively, of each input feature vector (defined to be the complex optical field). Consequently, including an LO generates a vector space with dimension $(\nin + 1)^2$. We find that an LO is not necessary for adequate channel equalization in this work, likely because the IM/DD symbols were encoded into the optical power to begin with and subsequently distorted. As a result, being able to isolate the field terms provides no additional computational power compared to the ability to isolate the power (line 1 in \cref{eq:voutnl}). However, we still find it useful that  \cref{eq:voutnl,eq:voutlo} include information about the full complex field; even though information is encoded into the magnitude squared of the field, the optical-domain effects of chromatic dispersion and Kerr nonlinearity introduce distortions onto the complex field in a way that is not possible to compensate by computations applied to the optical power only.
	
	As a final note, we can compare the photonic output feature vectors to digital NGRCs as well as those detailed by \textcite{wangUltrafastSiliconPhotonic2024}. We found in this work that the underlying output feature vectors existed in a space of dimension $\nin^2$ (no LO) or $(\nin + 1)^2$ (with LO). We can make connection to the work in Ref. \cite{wangUltrafastSiliconPhotonic2024} by noting that their feature vector was encoded strictly into the real part of the optical field using a modulator biased at the null point. In the context of \cref{eq:voutnl,eq:voutlo}, we simply set $v_{\text{in},i}^I[l] = 0$ to recover a vector space of dimension $\nin(\nin + 1)/2$ (no LO) and $(\nin + 1)(\nin + 2)/2$ (with LO), since any quantity that includes a factor of $v_{\text{in},i}^I[l]$ is zero. In the case that initial data is encoded into the optical field, then the full-rank photonic NGRC with LO is equivalent to a second order NGRC, which is simply \cref{eq:ngrc3} reduced to a sum over all combinations of two input feature vector elements. However, this comparison does not apply in this work: the output feature vectors in \cref{eq:ngrc3} are formed from combinations of the optical power, so that even a second-order NGRC contains terms proportional to the fourth power of the optical field, whereas the photonic NGRC still contains only second-order field terms.

	\subsection{On-chip waveguide mixing}
	The simulated NGRC results in this work were achieved using a mixing region modeled by a random unitary matrix $\tilde{U}$ to generate output feature vectors. Here, we compare the NGRC performance for a PIC-based design which implements a matrix $\tilde{U}$ via mode mixing in a waveguide array structure. To model a photonic NGRC PIC with $\nout = 32$, we simulate a mixing region comprising 32 single-mode coupled waveguides. Each waveguide is chosen to be a silicon ridge with index $n_{\text{Si}} = 3.48$, width $400$ nm, and height $230$ nm, surrounded by SiO$_2$ with index $n_{\text{SiO}_2} = 1.44$. The waveguides are equally spaced by $300$ nm to ensure coupling between modes.
	
\begin{figure}[h!]
		\centering
		\includegraphics{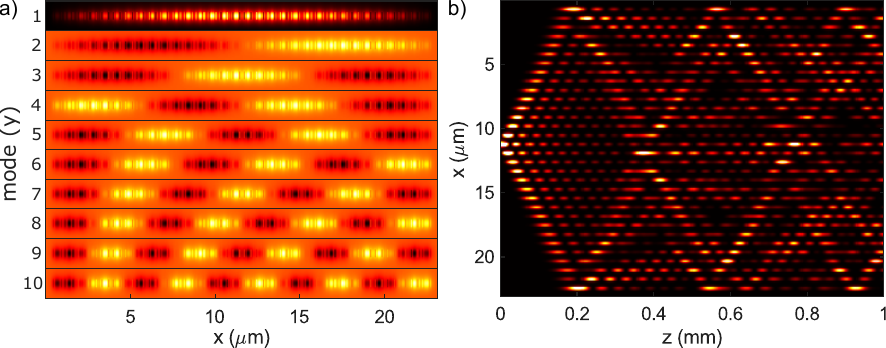}
		\caption{a) The x component of the electric field for the first 10 modes of the 32-waveguide mixing region. b) Propagation of the 16th input mode through the waveguide mixing region.}\label{fig:wg_prop}
	\end{figure}

	The effective index and electromagnetic fields of the first $\nin = 32$ modes of the waveguide array structure were computed for a 1550 nm wavelength by the finite difference frequency domain method using the Matlab Photonic FDFD community toolbox \cite{pangPhotonicFDFDToolbox2026}.  \Cref{fig:wg_prop}a shows the $x$ component of the electric field for the first ten of these array modes. To determine how these array modes are excited by incident single-mode waveguides, the system is re-simulated on the same mesh for a single waveguide located at each position and then the mode overlap between input waveguide $i$ and array mode $j$ is computed by
	\begin{equation}\label{eq:Cij}
		C_{ij} = \frac{1}{2}\text{Re}\left\{\mathbf{E}_i\times \mathbf{H}_j^*\right\},
	\end{equation}
	where all individual and array modes $\alpha$ are normalized to unit power by
	\begin{equation}
		P_\alpha = \int S_z(x,y)\,dxdy = \frac{1}{2} \int \text{Re}\left\{\mathbf{E}_\alpha\times \mathbf{H}_\alpha^*\right\}\cdot\hat{\mathbf{z}}\,dxdy = 1.
	\end{equation} 
	Using \cref{eq:Cij}, the effect of propagation is calculated
	\begin{equation}\label{eq:Uz}
		\tilde{U}(z) = C^\dagger e^{\beta k_0 z} C,
	\end{equation}
	where $k_0 = 2\pi/\lambda$ and $\beta$ is a diagonal matrix with entries $\beta_{ii} = n_{\text{eff},i}$ given by the effective index of mode $i$. Note that the resulting $\tilde{U}$ is only approximately physically correct because $C$ is not exactly unitary due to the non-orthogonality of the single-waveguide modes represented in the first index. However, we do not expect the equalization results to be heavily dependent on these small off-diagonal terms in $C^\dagger C$. \Cref{fig:wg_prop}b depicts the propagation of the optical irradiance integrated over the $y$ direction for the waveguide array excited by light coupled into the 16th waveguide. 
	
	To compare the effect of using a PIC-based mixing instead of a truly random unitary matrix, we repeat the simulations of \cref{fig:ber_2d_fixed_dimension} where $\nout = 32$, $M = 4$, and $\ros = 5$. Repeating the simulation for these hyperparameters five times with different random matrices, we find $\text{BER} = 5 \times 10^{-5}$. Then, performing a noise-free simulation of the NGRC according to \cref{eq:Amix} with $\tilde{U}$ given by \cref{eq:Uz} with $z = L$, we achieve an error rate of $\text{BER} = 5 \times 10^{-5}$ for the simulated waveguide system, which is the same as what we found for random mixing. Since the error rate remained well below the KP4 threshold, we conclude that this coupled waveguide mixing is satisfactory for PAM4 channel equalization.

\clearpage
\bibliography{MyLibrary}
\end{document}